\documentclass{elsart}
\usepackage{amsmath,amssymb,graphicx,subfigure,url}
\DeclareGraphicsExtensions{.eps}
\journal{Physica A}
\begin{document}


\begin{frontmatter}
\title{The $\kappa$-generalized distribution: A new descriptive model for the size distribution of incomes}
\author[Ancona]{F. Clementi\corauthref{cor}},
\corauth[cor]{Corresponding author: Tel.: +39--071--22--07--103; fax: +39--071--22--07--102.}
\ead{fabio.clementi@univpm.it}
\author[Canberra]{T. Di Matteo},
\ead{tiziana.dimatteo@anu.edu.au}
\author[Ancona]{M. Gallegati},
\ead{mauro.gallegati@univpm.it}
\author[Torino]{G. Kaniadakis}
\ead{giorgio.kaniadakis@polito.it}
\address[Ancona]{Department of Economics, Polytechnic University of Marche, Piazzale R. Martelli 8, 60121 Ancona, Italy}
\address[Canberra]{Applied Mathematics, Research School of Physical Sciences and Engineering, The Australian National University, 0200 Canberra, Australia}
\address[Torino]{Department of Physics, Polytechnic University of Turin, Corso Duca degli Abruzzi 24, 10129 Torino, Italy}
\begin{abstract}
This paper proposes the $\kappa$-generalized distribution as a model for describing the distribution and dispersion of income within a population. Formulas for the shape, moments and standard tools for inequality measurement\textemdash such as the Lorenz curve and the Gini coefficient\textemdash are given. A method for parameter estimation is also discussed. The model is shown to fit extremely well the data on personal income distribution in Australia and the United States.
\end{abstract}
\begin{keyword}
Income distribution\sep income inequality\sep $\kappa$-generalized statistics
\PACS 02.50.Ng, 02.60.Ed, 89.65.Gh
\end{keyword}
\end{frontmatter}


\section{Introduction}
\label{sec:Introduction}
In the analysis of income distributions, analysts have found it useful to have distributional summaries based on estimates of specific parametric functional forms, not only for their suitability in modelling some features of many empirical income distributions, but also because of their role as equilibrium distributions in economic processes \cite{RichmondHutzlerCoelhoRepetowicz2006}.
\par
Vilfredo Pareto first proposed a model of income distribution in the form of a probability density function in 1897 \cite{Pareto1897}, providing a description of the density for income values above some lower bound, $x_{0}>0$ \cite{Arnold1983}. If one focuses on the distribution amongst those with income greater than $x_{0}$, there are simple expressions for the moments which depend only on the Pareto parameters $\alpha$ and $x_{0}$. Moreover, the expressions for most common inequality measures depend only on $\alpha$, so that the (inverse of) $\alpha$ may also be considered as an inequality measure.
\par
However, the apparent attractions of the Pareto distribution evaporate somewhat when one considers its implications for the distribution of income amongst the population as a whole, \textit{i.e.} including units with income less than $x_{0}$. For example, Ref. \cite{AtkinsonHarrison1978} shows how the expression for the Gini coefficient depends on assumptions about the size of excluded population (\textit{i.e.} the proportion of the population with income below $x_{0}$) and its average income. In particular, $\alpha$ no longer has such a straightforward interpretation. For example, an increase in $\alpha$ may be associated with a decrease in inequality according to the Gini coefficient, but an increase according to the coefficient of variation.
\par
Later empirical studies showed that the income distribution is right-skewed and has a fat right-hand tail, and that the Pareto distribution accurately models only high levels of income, but does a poor job in describing the lower end of the distribution. As research continued, new models were proposed that better describe the data, using both a combination of known statistical distributions \cite{Combination} and parametric functional forms for the distribution of income as a whole, including two-parameter models such as the lognormal and gamma, three-parameter distributions such as the Singh-Maddala and Dagum I, and four-parameter distributions such as the generalized beta distributions of the first and second kind (see the comprehensive survey by Ref. \cite{KleiberKotz2003}).
\par
The Pareto fat tails was observed experimentally also in physical statistical systems, and are located in the high energy region where the laws of classical physics are replaced by the relativistic ones. After 2001, a physical mechanism emerging in the context of special relativity was proposed by one of us \cite{Kaniadakis}, predicting a deformation of the exponential function. According to this mechanism, the classical exponential distribution transforms into  a new distribution, which at high energies presents a Pareto fat tail. More precisely, this mechanism deforms the ordinary exponential function $\exp\left(x\right)$ into the generalized exponential function $\exp_{\kappa}\left(x\right)$ given by
\begin{equation}
\exp_{\kappa}\left(x\right)=\left(\sqrt{1+\kappa^{2}x^{2}}+\kappa x\right)^{\frac{1}{\kappa}}.
\label{eq:Equation1}
\end{equation}
The above deformation is generated by the fact that the propagation of the information has a finite speed, and the deformation parameter $\kappa$ is proportional to the reciprocal of this speed. The $\kappa$-generalized exponential has the important properties
\begin{subequations}
\begin{equation}
\exp_{\kappa}\left(x\right){\atop\stackrel{\textstyle\sim}{\scriptstyle x\rightarrow\pm\infty}}\left|2\kappa x\right|^{\pm\frac{1}{\left|\kappa\right|}},
\label{eq:Equation2a}
\end{equation}
\begin{equation}
\exp_{\kappa}\left(x\right){\atop\stackrel{\textstyle\sim}{\scriptstyle x\rightarrow\,0}}\exp\left(x\right).
\label{eq:Equation2b}
\end{equation}
\label{eq:Equation2}
\end{subequations}
\par
It is remarkable that for classical systems where the information propagates instantaneously it results $\kappa=0$, so that the ordinary exponential emerges naturally after noting that $\exp_{0}\left(x\right)=\exp\left(x\right)$. Moreover, in the low energy region $x\rightarrow0$ according to Eq. \eqref{eq:Equation2b} the exponential distribution emerges again, because the system behaves classically. On the contrary, in systems where the information propagates with a finite speed\textemdash these systems are intrinsically relativistic\textemdash it results $\kappa\neq0$, so that the exponential tails become fat according to Eq. \eqref{eq:Equation2a} and the Pareto law emerges.
\par
The generalized exponential represents a very useful and powerful tool to formulate a new statistical theory capable to treat systems described by distribution functions exhibiting power-law tails and admitting a stable entropy \cite{KaniadakisScarfone,AbeKaniadakisScarfone}. Furthermore, non-linear evolution models already known in statistical physics \cite{KaniadakisLavagnoQuarati} can be easily adapted or generalized within the new theory.
\par
After 2001, the function  $\exp_{\kappa}\left(x\right)$ was adopted successfully in  the analysis of various physical and non physical systems. In Ref. \cite{ClementiGallegatiKaniadakis2007} we have used the function $\exp_{\kappa}\left(x\right)$ to model the personal income distribution by defining the Complementary Cumulative Distribution Function (CCDF) through
\begin{equation}
P_{>}\left(x\right)=\exp_{\kappa}\left(-\beta x^{\alpha}\right),\quad x\in\mathbf{R}^{+},\quad\alpha,\beta>0,\quad\kappa\in[0,1),
\label{eq:Equation3}
\end{equation}
where the income variable $x$ is defined as $x=\frac{z}{\left\langle z\right\rangle}$, being $z$ the absolute personal income and $\left\langle z\right\rangle$ its mean value. The corresponding Probability Density Function (PDF) reads
\begin{equation}
p\left(x\right)=\frac{\alpha\beta x^{\alpha-1}\exp_{\kappa}\left(-\beta x^{\alpha}\right)}{\sqrt{1+\kappa^2\beta^2x^{2\alpha}}}.
\label{eq:Equation4}
\end{equation}
Follows immediately that in this model for low incomes the CCDF behaves as a stretched exponential $P_{>}\left(x\right)=\exp\left(-\beta x^{\alpha}\right)$, while at high incomes follows the Pareto law $P_>\left(x\right)=\left(2\beta\kappa\right)^{-\frac{1}{\kappa}}x^{-\frac{\alpha}{\kappa}}$. Similarly, the PDF for $x\rightarrow0^{+}$ behaves as a Weibull distribution $p\left(x\right)=\alpha\beta x^{\alpha-1}\exp\left(-\beta x^{\alpha}\right)$, while for $x\rightarrow+\infty$ reduces to the Pareto's law
$p\left(x\right)=\frac{\alpha}{\kappa}\left(2\beta\kappa\right)^{-\frac{1}{\kappa}}x^{-\left(\frac{\alpha}{\kappa}+1\right)}$.
\par
Starting from the definitions in Eqs. \eqref{eq:Equation3} and \eqref{eq:Equation4}, in this work we derive the basic statistical properties of the proposed distribution along with common tools that are required for income distribution analysis; these include, among others, the ubiquitous Lorenz curve and the associated Gini measure of inequality. The basic proposition of this paper is that the $\kappa$-distribution provides a very good description of the size distribution of income, ranging from the low region to the middle region, and up to the power-law tail, and the inequality analysis expressed in terms of its parameters reveals very powerful.
\par
The content of the paper is organized as follows: Sec. \ref{sec:TheKappaGeneralizedDistribution} includes a discussion of the $\kappa$-distribution and reports formulas which are useful in the estimation and analysis of empirical data. Sec. \ref{sec:EmpiricalApplicationToIncomeData} illustrates applications of the results to Australian and US household survey data. Sec. \ref{sec:SummaryAndConclusions} concludes.


\section{The $\kappa$-generalized distribution}
\label{sec:TheKappaGeneralizedDistribution}


\subsection{Basic properties}
\label{sec:BasicProperties}
\par
Using the complementary relation $P_{\leq}\left(x\right)=1-P_{>}\left(x\right)$, we see that the quantile function is available in closed form
\begin{equation}
x=P^{-1}_{\leq}\left(u\right)=\beta^{-\frac{1}{\alpha}}\left[\log_{\kappa}\left(\frac{1}{1-u}\right)\right]^{\frac{1}{\alpha}},\quad0<u<1,
\label{eq:Equation5}
\end{equation}
a property that facilitates the derivation of Lorenz-ordering results (see Sec. \ref{sec:LorenzCurvesAndInequalityMeasures}). From Eq. \eqref{eq:Equation5} we easily determine that the median of the distribution is $x_{\mathrm{med}}=\beta^{-\frac{1}{\alpha}}\left[\log_{\kappa}\left(2\right)\right]^{\frac{1}{\alpha}}$.
\par
The mode is at
\begin{equation}
x_{\mathrm{mode}}\!=\!\beta^{-\frac{1}{\alpha}}\left\{\left[\frac{\alpha+2\kappa^{2}\left(\alpha-1\right)}{2\kappa^{2}\left(\alpha^{2}-\kappa^{2}\right)}\right]\left(\sqrt{1+\frac{4\kappa^{2}\left(\alpha^{2}-\kappa^{2}\right)\left(\alpha-1\right)^{2}}{\left[\alpha^{2}+2\kappa^{2}\left(\alpha-1\right)\right]^{2}}}-1\right)\right\}^{\frac{1}{2\alpha}}
\label{eq:Equation6}
\end{equation}
if $\alpha>1$; otherwise, the distribution is zero-modal with a pole at the origin.


\subsection{Moments and related parameters}
\label{sec:MomentsAndRelatedParameters}
The $r$\textsuperscript{th}-order moment about the origin of the $\kappa$-generalized distribution equals
\begin{equation}
\mu^{'}_{r}=\int\limits_{0}^{\infty}x^{r}p\left(x\right)\operatorname{d}x=\frac{\left(2\beta\kappa\right)^{-\frac{r}{\alpha}}}{1+\frac{r}{\alpha}\kappa}\frac{\Gamma\left(\frac{1}{2\kappa}-\frac{r}{2\alpha}\right)}{\Gamma\left(\frac{1}{2\kappa}+\frac{r}{2\alpha}\right)}\Gamma\left(1+\frac{r}{\alpha}\right),
\label{eq:Equation7}
\end{equation}
where $\Gamma\left(x\right)$ is the Gamma function $\Gamma\left(x\right)=\int_{0}^{\infty}t^{x-1}e^{-t}\mathrm{d}t$. Specifically, $\mu^{'}_{1}=m$ is the mean of the distribution.
\par
A formula for the variance is obtained by converting Eq. \eqref{eq:Equation7} to the moment about the mean using the general equation $\mu_{r}=\sum^{r}_{j=0}{r\choose j}\left(-1\right)^{r-j}\mu^{'}_{j}m^{r-j}$; hence, for $r=2$ we have
\begin{equation}
\sigma^{2}=\left(2\beta\kappa\right)^{-\frac{2}{\alpha}}\left\{\frac{\Gamma\left(1+\frac{2}{\alpha}\right)}{1+2\frac{\kappa}{\alpha}}\frac{\Gamma\left(\frac{1}{2\kappa}-\frac{1}{\alpha}\right)}{\Gamma\left(\frac{1}{2\kappa}+\frac{1}{\alpha}\right)}-\left[\frac{\Gamma\left(1+\frac{1}{\alpha}\right)}{1+\frac{\kappa}{\alpha}}\frac{\Gamma\left(\frac{1}{2\kappa}-\frac{1}{2\alpha}\right)}{\Gamma\left(\frac{1}{2\kappa}+\frac{1}{2\alpha}\right)}\right]^{2}\right\}.
\label{eq:Equation8}
\end{equation}
\par
In this way it is also possible to define the standardized moments of the distribution, which are in turn used to define skewness and excess kurtosis, respectively given by
\begin{equation}
\gamma_{1}=\frac{\mu_{3}}{\sigma^{3}}=\frac{\mu^{'}_{3}-3m\sigma^{2}-m^{3}}{\sigma^{3}}
\label{eq:Equation9}
\end{equation}
and
\begin{equation}
\gamma_{2}=\frac{\mu_{4}}{\sigma^{4}}-3=\frac{\mu^{'}_{4}-3\sigma^{4}-4\gamma_{1}\sigma^{3}m-6\sigma^{2}m^{2}-m^{4}}{\sigma^{4}}.
\label{eq:Equation10}
\end{equation}


\subsection{Lorenz curves and inequality measures}
\label{sec:LorenzCurvesAndInequalityMeasures}
For a discussion of income inequality, the standard practice adopts the concept of concentration of incomes as defined by Lorenz \cite{Lorenz1905}. The so-called Lorenz curve measures the cumulative fraction of population with incomes below $x$ along the horizontal axis, and the fraction of the total income this population accounts for along the vertical axis. The points plotted for the various values of $x$ trace out a curve below the $45^{\circ}$ line sloping upwards to the right from the origin.
\par
In statistical terms, for any general distribution supported on the nonnegative half-line with a finite and positive first moment the Lorenz curve can be written in the form $L\left(u\right)=\int^{u}_{0}P_{\leq}^{-1}\left(t\right)\operatorname{d}t/\int^{1}_{0}P_{\leq}^{-1}\left(u\right)\operatorname{d}u$, $u\in\left[0,1\right]$, where $m=\int^{1}_{0}P_{\leq}^{-1}\left(u\right)\operatorname{d}u$ is the quantile formula for the mean, and $P_{\leq}^{-1}\left(u\right)$ is the quantile function given by Eq. \eqref{eq:Equation5}\footnote{See \textit{e.g.} Ref. \cite{Gastwirth1971}.}. Thus, we have the Lorenz curve for the $\kappa$-generalized distribution as follows
\begin{equation}
\begin{split}
L_{\kappa}\left(u\right)=&1-\frac{1+\frac{\kappa}{\alpha}}{2\Gamma\left(\frac{1}{\alpha}\right)}\frac{\Gamma\left(\frac{1}{2\kappa}+\frac{1}{2\alpha}\right)}{\Gamma\left(\frac{1}{2\kappa}-\frac{1}{2\alpha}\right)}\left\{2\alpha\left(2\kappa\right)^{\frac{1}{\alpha}}\left(1-u\right)\left[\log_{\kappa}\left(\frac{1}{1-u}\right)\right]^{\frac{1}{\alpha}}+\right.\\
&\Biggl.+B_{X}\left(\frac{1}{2\kappa}-\frac{1}{2\alpha},\frac{1}{\alpha}\right)+B_{X}\left(\frac{1}{2\kappa}-\frac{1}{2\alpha}+1,\frac{1}{\alpha}\right)\Biggr\},
\end{split}
\label{eq:Equation11}
\end{equation}
\sloppy where $B_{x}\left(s,r\right)$ is the incomplete Beta function given by $B_{x}\left(s,r\right)=\int_{0}^{x}t^{s-1}\left(1-t\right)^{r-1}\mathrm{d}t$ with $X=\left(1-u\right)^{2\kappa}$.
\par
\fussy The Gini measure of income inequality \cite{Gini1914} can be derived using the representation $G=1-\frac{1}{m}\int^{\infty}_{0}\left[1-P_{\leq}\left(x\right)\right]^{2}\operatorname{d}x$ given by Ref. \cite{ArnoldLaguna1977}; it follows that the Gini coefficient for the $\kappa$-generalized distribution is
\begin{equation}
G_{\kappa}=1-\frac{2\alpha+2\kappa}{2\alpha+\kappa}\frac{\Gamma\left(\frac{1}{\kappa}-\frac{1}{2\alpha}\right)}{\Gamma\left(\frac{1}{\kappa}+\frac{1}{2\alpha}\right)}\frac{\Gamma\left(\frac{1}{2\kappa}+\frac{1}{2\alpha}\right)}{\Gamma\left(\frac{1}{2\kappa}-\frac{1}{2\alpha}\right)}.
\label{eq:Equation12}
\end{equation}
\par
Furthermore, relating the standard deviation to the mean yields the following expression for the coefficient of variation
\begin{equation}
CV_{\kappa}=\frac{\sigma}{m}=\sqrt{\frac{\frac{\Gamma\left(1+\frac{2}{\alpha}\right)}{1+2\frac{\kappa}{\alpha}}\frac{\Gamma\left(\frac{1}{2\kappa}-\frac{1}{\alpha}\right)}{\Gamma\left(\frac{1}{2\kappa}+\frac{1}{\alpha}\right)}}{\left[\frac{\Gamma\left(1+\frac{1}{\alpha}\right)}{1+\frac{\kappa}{\alpha}}\frac{\Gamma\left(\frac{1}{2\kappa}-\frac{1}{2\alpha}\right)}{\Gamma\left(\frac{1}{2\kappa}+\frac{1}{2\alpha}\right)}\right]^{2}}-1}.
\label{eq:Equation13}
\end{equation}


\subsection{Estimation}
\label{sec:Estimation}
Parameter estimation for the $\kappa$-generalized distribution can be performed using the Maximum Likelihood (ML) approach. Assuming that all observations $\mathbf{x}=\left\{x_{1},\ldots,x_{n}\right\}$ are independent, the likelihood function is
\begin{equation}
L\left(\boldsymbol{\theta};\mathbf{x}\right)=\prod\limits^{n}_{i=1}p\left(x_{i}\right)=\left(\alpha\beta\right)^{n}\prod\limits^{n}_{i=1}\frac{x_{i}^{\alpha-1}\exp_{\kappa}\left(-\beta x_{i}^{\alpha}\right)}{\sqrt{1+\beta^{2}\kappa^{2}x_{i}^{2\alpha}}},
\label{eq:Equation14}
\end{equation}
where $\boldsymbol{\theta}=\left\{\alpha,\beta,\kappa\right\}$ is the parameter vector. This leads to the problem of solving the partial derivatives of the log-likelihood function $l\left(\boldsymbol{\theta};\mathbf{x}\right)=\log L\left(\boldsymbol{\theta};\mathbf{x}\right)$ with respect to $\kappa$, $\alpha$ and $\beta$. However, obtaining explicit expressions for the ML estimators of the three parameters is difficult, making direct analytical solutions intractable, and one needs to use numerical optimization methods.
\par
Taking into account the meaning of the variable $x$, the mean value results to be equal to unity, \textit{i.e.} $m=\int_{0}^{\infty}xp\left(x\right)\mathrm{d}x=1$. The latter relationship permits to express the parameter $\beta$ as a function of the parameters $\kappa$ and $\alpha$, obtaining
\begin{equation}
\beta=\frac{1}{2\kappa}\left[\frac{\Gamma\left(\frac{1}{\alpha}\right)}{\kappa+\alpha}\frac{\Gamma\left(\frac{1}{2\kappa}-\frac{1}{2\alpha}\right)}{\Gamma\left(\frac{1}{2\kappa}+\frac{1}{2\alpha}\right)}\right]^{\alpha}.
\label{eq:Equation15}
\end{equation}
In this way, the problem to determine the values of the free parameters $\left\{\kappa,\alpha,\beta\right\}$ of the theory from the empirical data reduces to a two parameter $\left\{\kappa,\alpha\right\}$ fitting problem. Therefore, to find the parameter values that give the most desirable fit, one can use the Constrained Maximum Likelihood (CML) estimation method \cite{Schoenberg1997}, which solves the general maximum log-likelihood problem of the form $l\left(\boldsymbol{\theta};\mathbf{x}\right)=\sum^{n}_{i=1}\log p\left(x_{i};\boldsymbol{\theta}\right)^{w_{i}}$, where $n$ is the number of observations, $w_{i}$ the weight assigned to each observation, $p\left(x_{i};\boldsymbol{\theta}\right)$ the probability of $x_{i}$ given $\boldsymbol{\theta}$, subject to the non-linear equality constraint given by Eq. \eqref{eq:Equation15} and bounds $\alpha,\beta>0$ and $\kappa\in\left[0,1\right)$. The CML procedure finds values for the parameters in $\boldsymbol{\theta}$ such that $l\left(\boldsymbol{\theta};\mathbf{x}\right)$ is maximized using the sequential quadratic programming method \cite{Han1977} as implemented, \textit{e.g.}, in \textsc{Matlab}\textsuperscript{\textregistered} 7.


\setcounter{equation}{0}
\section{Empirical application to income data}
\label{sec:EmpiricalApplicationToIncomeData}
The $\kappa$-generalized distribution was fit to data on personal income distribution for Australia and the United States\footnote{These data were not studied in the previous paper \cite{ClementiGallegatiKaniadakis2007}, where the emphasis was on other countries. However, our main findings here have been applied also to the data included in that work, and the results are available from the authors upon request.}. The data are derived from panel surveys conducted in 2002--03 and 2003, respectively. The unit of assessment is the household, and income is expressed in nominal local currency units (and is equivalized for differences in household size by adjusting by the square root of the number of household members \cite{Deaton1996}). There are 10,211 households in the 2002--03 Australian survey, and 7,822 in the 2003 US survey. All calculations use the sampling weights produced by the data provider \cite{TheCanberraGroup2001}. We consider the distributions of disposable income, \textit{i.e.} the income recorded after the payment of taxes and government transfers. In the data analysis, we exclude the observations with zero and negative values, and normalize income to its empirical average, given by $32,891.17\pm343.58$ AUD and $31,812.39\pm598.74$ USD respectively\footnote{More detailed information on the Australian data is available on the Australian Bureau of Statistics (ABS) web site: \url{http://www.abs.gov.au}. For analyses referring to the same country and data source, see Refs. \cite{Australia}. For the US data, see Refs. \cite{BurkhauserEtAl}, or consult the following web address: \url{http://www.human.cornell.edu/che/PAM/Research/Centers-Programs/German-Panel/cnef.cfm}.}.
\par
\begin{figure}[!t]
\centering
\begin{center}
\mbox{
\subfigure[\hspace{0.2cm}Complementary CDF]{\label{fig:Figure1_a}\includegraphics[width=0.48\textwidth]{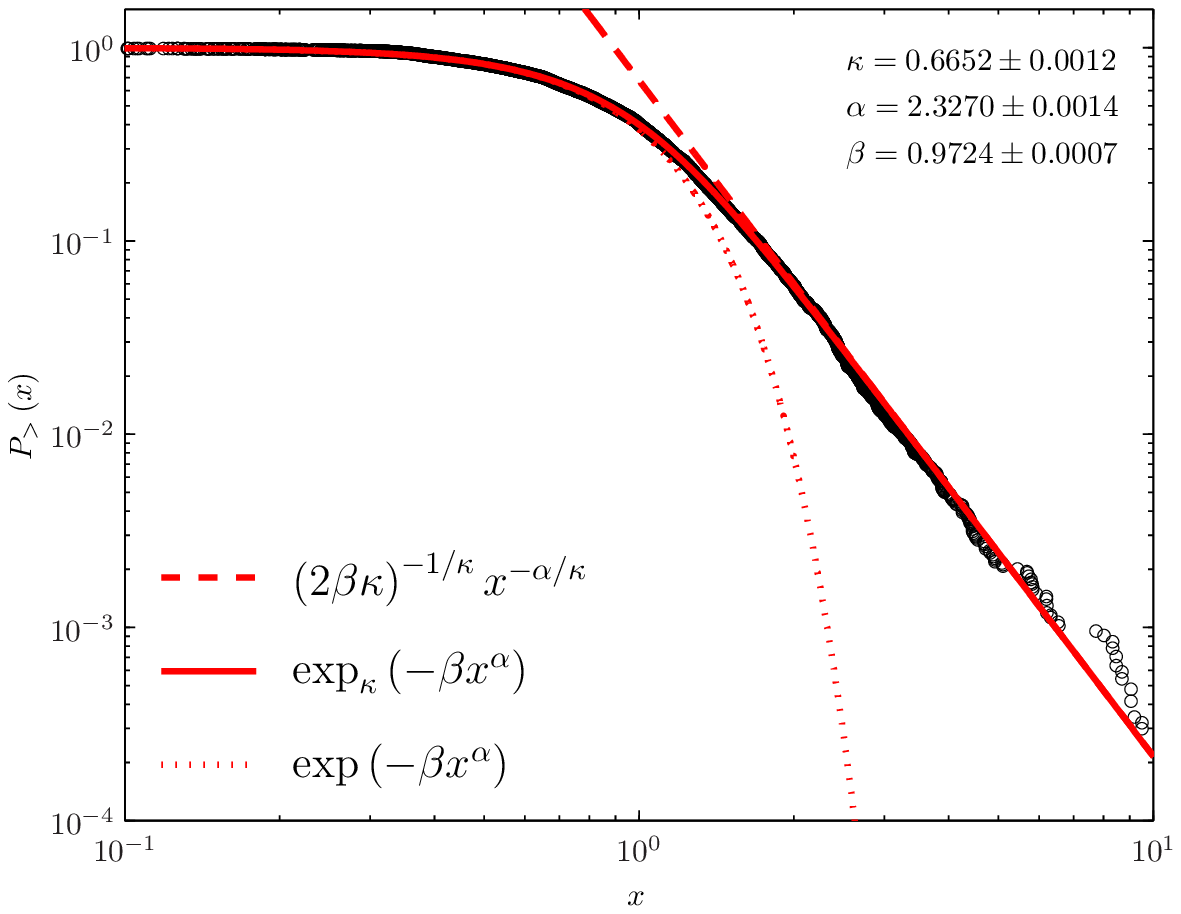}}
\subfigure[\hspace{0.2cm}PDF histogram plot]{\label{fig:Figure1_b}\includegraphics[width=0.48\textwidth]{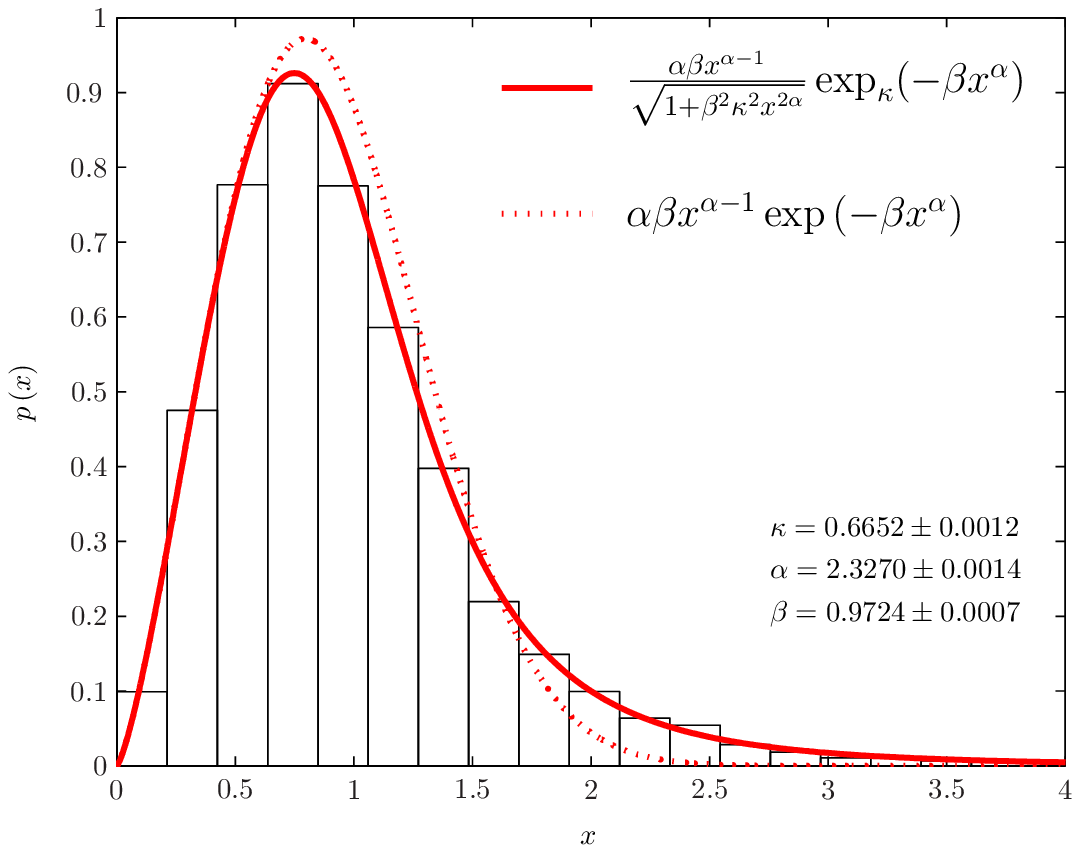}}
}
\mbox{
\subfigure[\hspace{0.2cm}Lorenz curve]{\label{fig:Figure1_c}\includegraphics[width=0.48\textwidth]{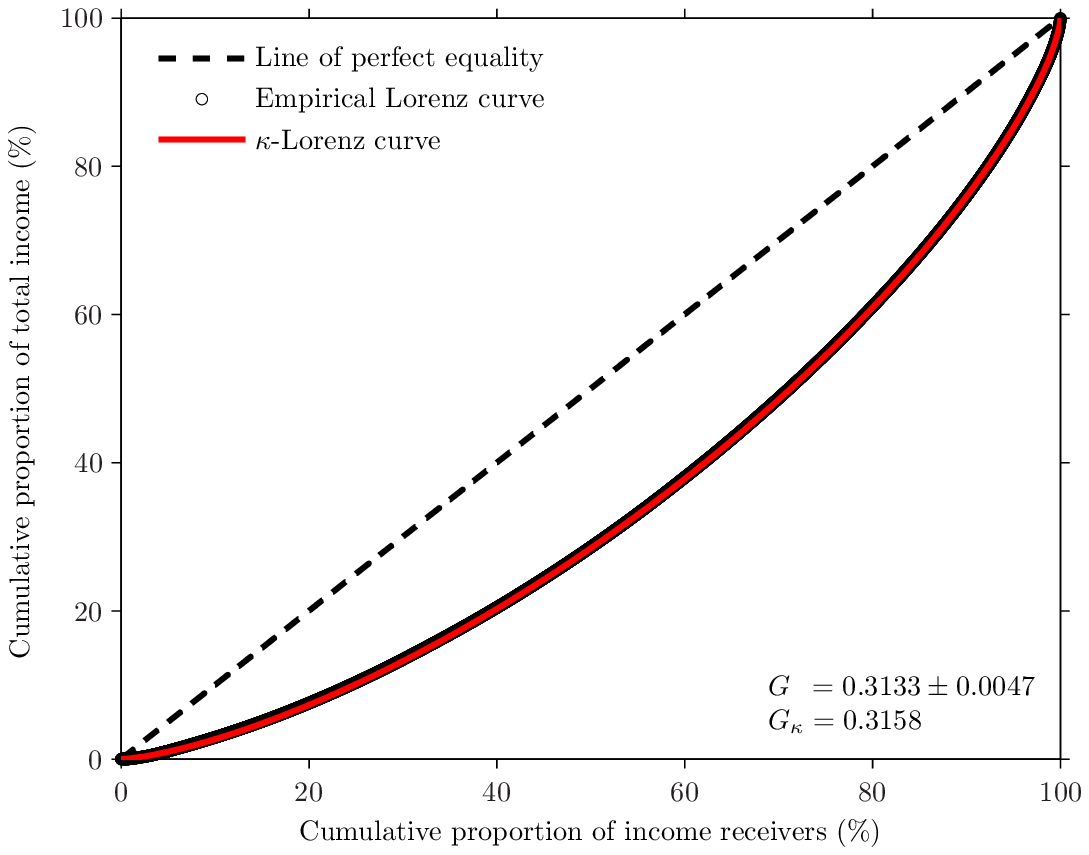}}
\subfigure[\hspace{0.2cm}Q-Q plot]{\label{fig:Figure1_d}\includegraphics[width=0.48\textwidth]{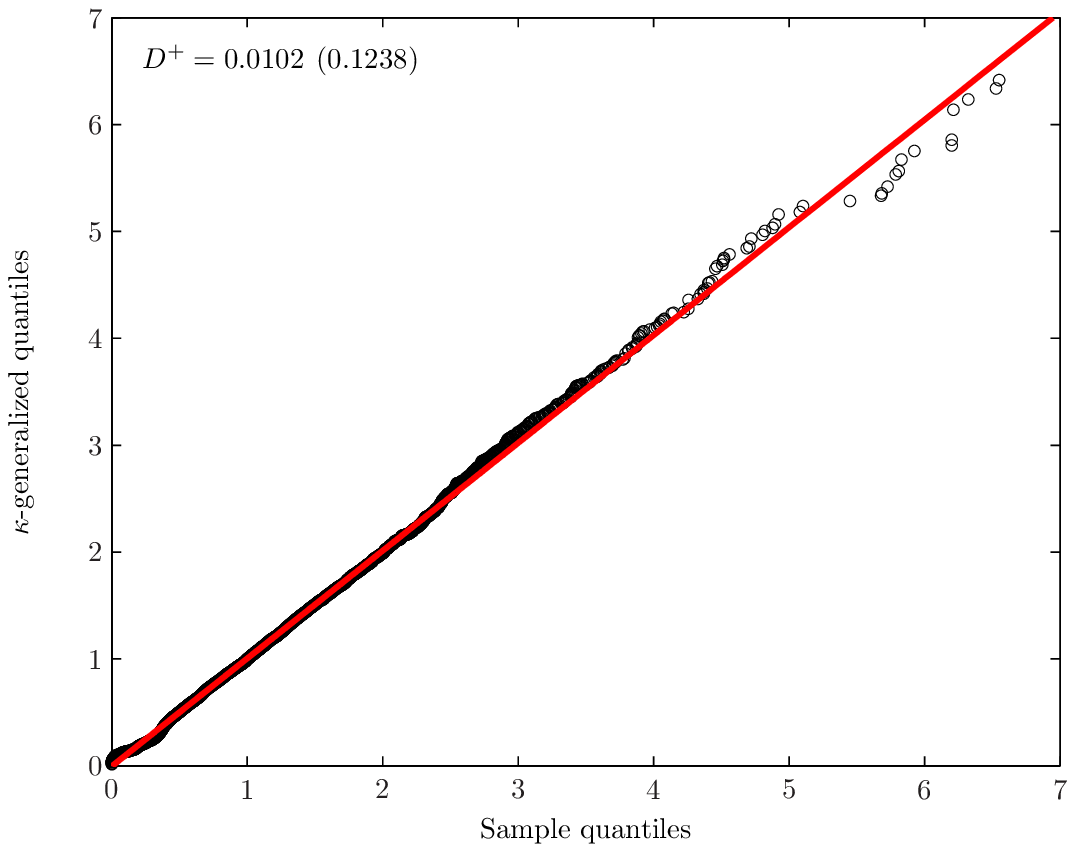}}
}
\caption{The Australian personal income distribution in 2002--03 measured in current year AUD. \subref{fig:Figure1_a} Plot of the empirical CCDF in the log-log scale. The solid line is our theoretical model given by Eq. \eqref{eq:Equation3} fitting very well the data in the whole range from the low to the high incomes including the intermediate income region. This function is compared with the ordinary stretched exponential one (dotted line)\textemdash fitting the low income data\textemdash and with the pure power-law (dashed line)\textemdash fitting the high income data. \subref{fig:Figure1_b} Histogram plot of the empirical PDF with superimposed fits of the $\kappa$-generalized (solid line) and Weibull (dotted line) PDFs. \subref{fig:Figure1_c} Plot of the Lorenz curve. The hollow circles represent the empirical data points and the solid line is the theoretical curve given by Eq. \eqref{eq:Equation11} using the same parameter values as in panels \subref{fig:Figure1_a} and \subref{fig:Figure1_b}. The dashed line corresponds to the Lorenz curve of a society in which everybody receives the same income and thus serves as a benchmark case against which actual income distribution may be measured. \subref{fig:Figure1_d} Q-Q plot of the sample quantiles versus the corresponding quantiles of the fitted $\kappa$-generalized distribution. The reference line has been obtained by locating points on the plot corresponding to around the 25\textsuperscript{th} and 75\textsuperscript{th} percentiles and connecting these two. In plots  \subref{fig:Figure1_a},  \subref{fig:Figure1_b} and  \subref{fig:Figure1_d} the income axis limits have been adjusted according to the range of data to shed light on the intermediate region between the bulk and the tail of the distribution.}
\label{fig:Figure1}
\end{center}
\end{figure}
\begin{figure}[!t]
\centering
\begin{center}
\mbox{
\subfigure[\hspace{0.2cm}Complementary CDF]{\label{fig:Figure2_a}\includegraphics[width=0.48\textwidth]{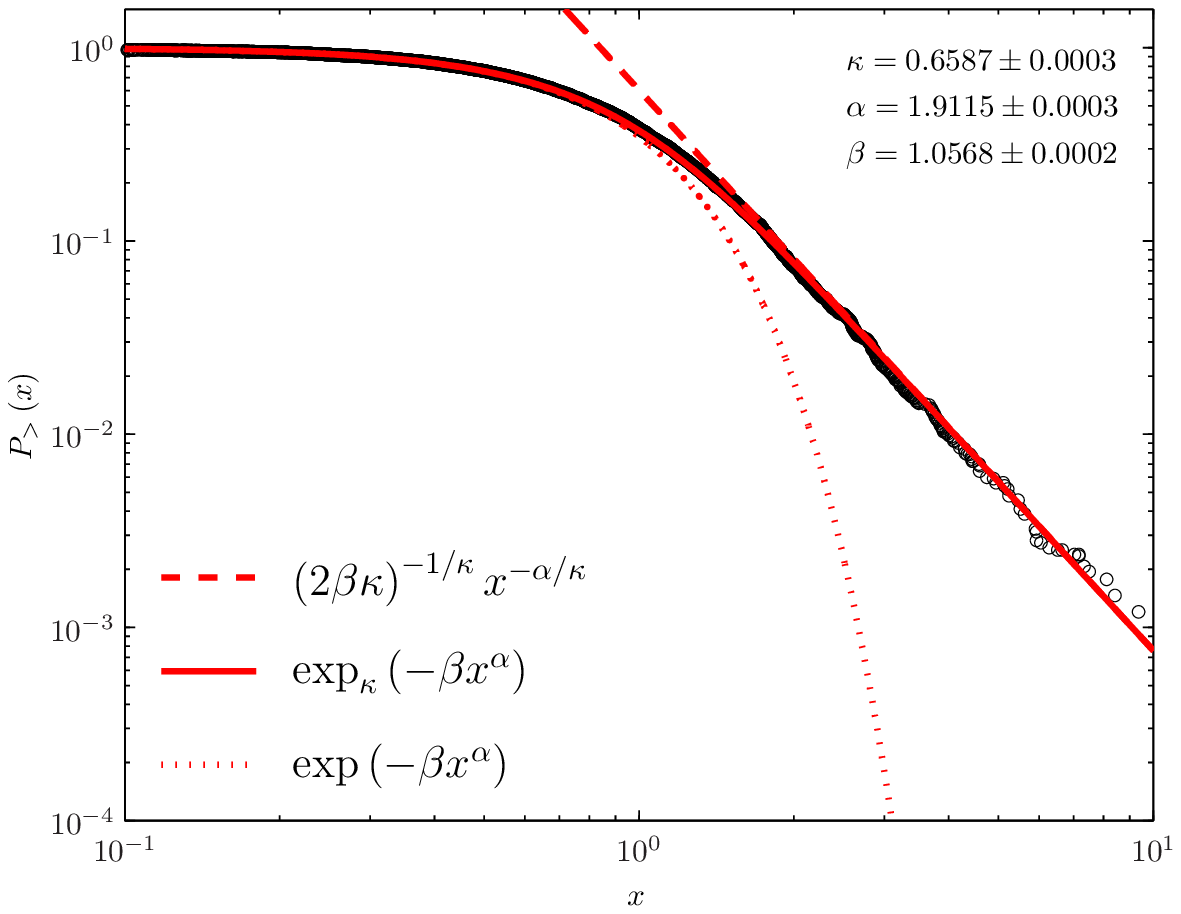}}
\subfigure[\hspace{0.2cm}PDF histogram plot]{\label{fig:Figure2_b}\includegraphics[width=0.48\textwidth]{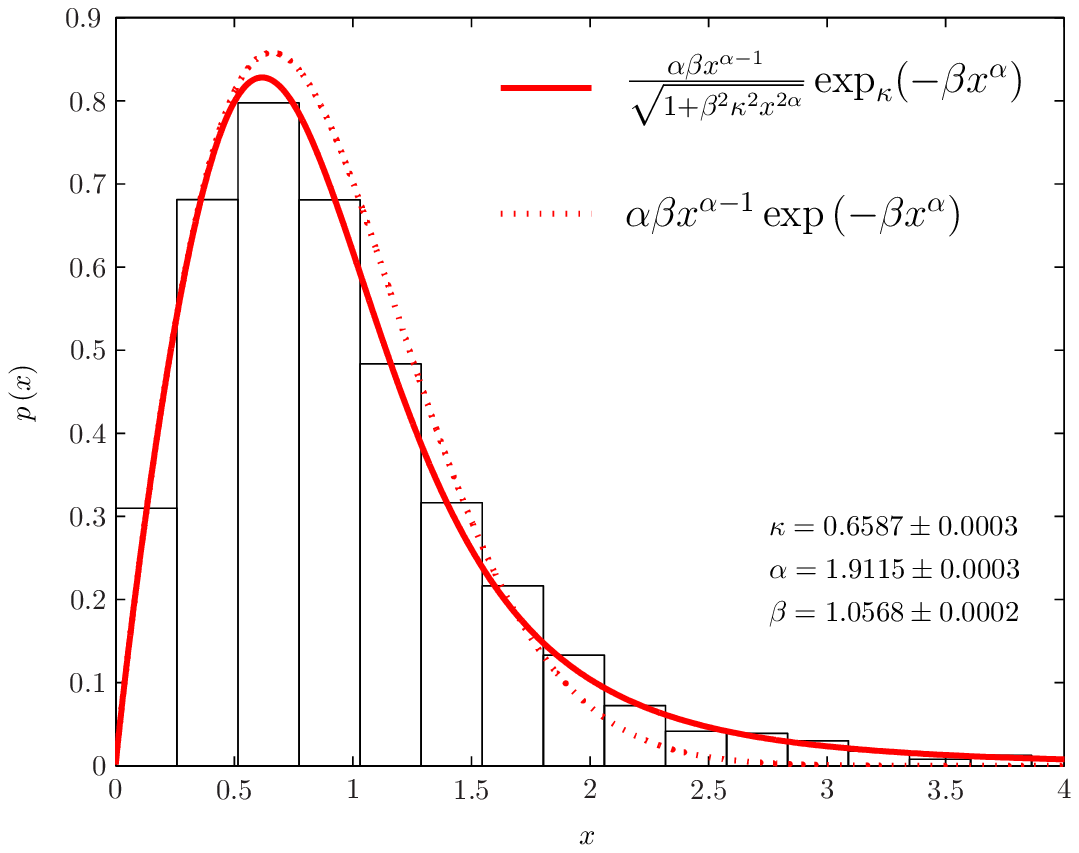}}
}
\mbox{
\subfigure[\hspace{0.2cm}Lorenz curve]{\label{fig:Figure2_c}\includegraphics[width=0.48\textwidth]{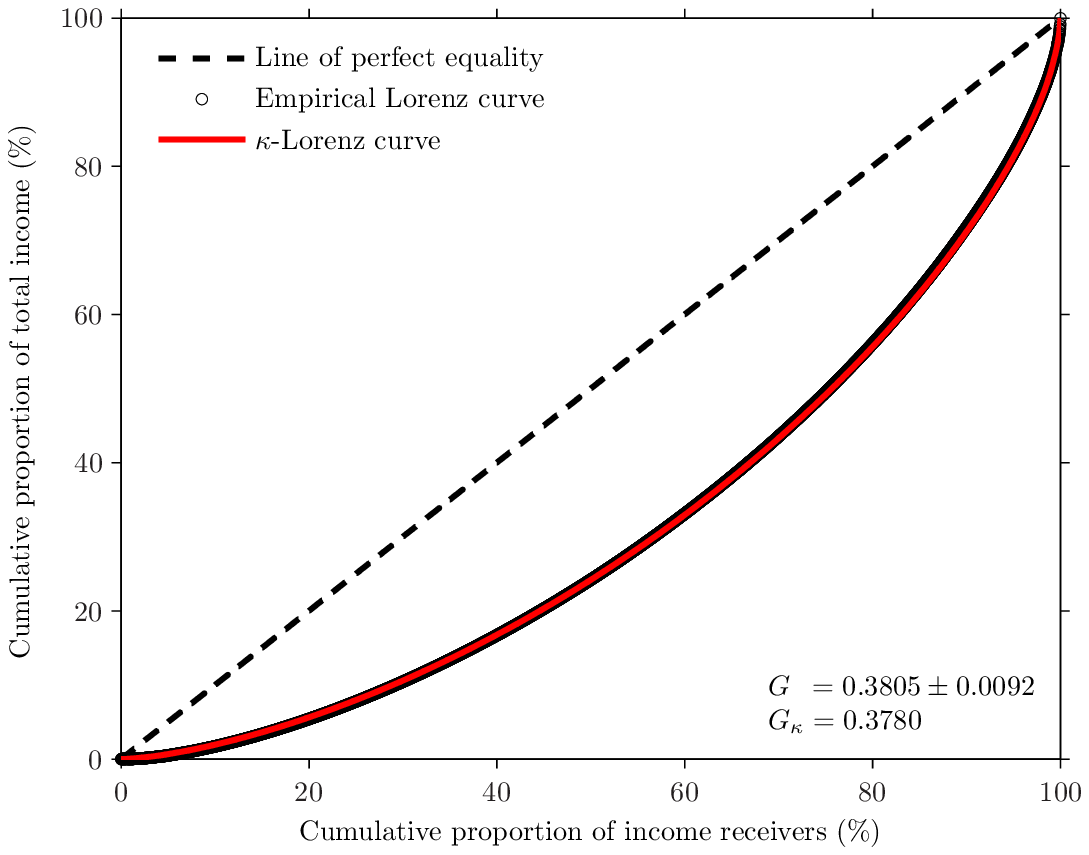}}
\subfigure[\hspace{0.2cm}Q-Q plot]{\label{fig:Figure2_d}\includegraphics[width=0.48\textwidth]{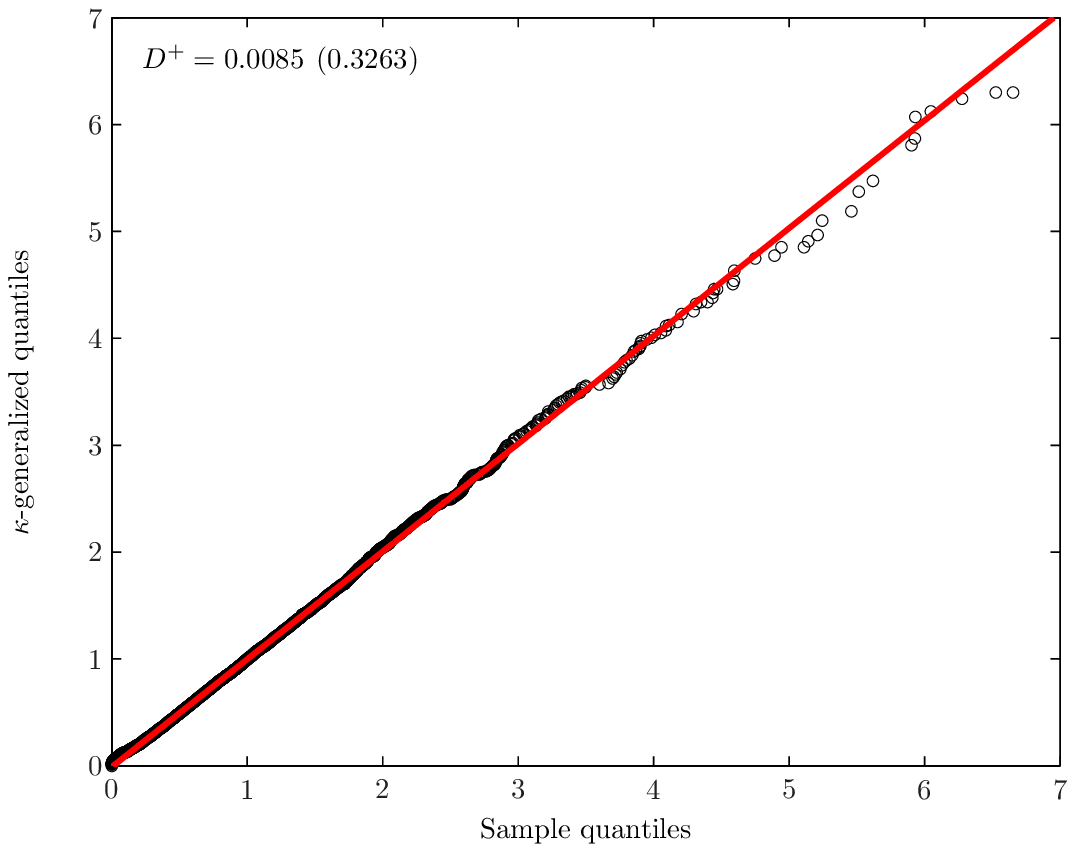}}
}
\caption{Same plots as in Fig. \ref{fig:Figure1} for the US personal income distribution in 2003. The income variable is measured in current year USD.}
\label{fig:Figure2}
\end{center}
\end{figure}

Maximum likelihood estimates are shown in panels \subref{fig:Figure1_a} and \subref{fig:Figure1_b} of Figs. \ref{fig:Figure1} and \ref{fig:Figure2}. All the parameters were very precisely estimated, and the comparison between the fitted and sample estimates of the CCDF and PDF suggests that the $\kappa$-generalized distribution offers a great potential for describing the data over their whole range, from the low to medium income region through to the high income Pareto power-law regime, including the intermediate region for which a clear deviation exists when two different curves are used.
\par
Panel \subref{fig:Figure1_c} of the same figures depicts the data points for the empirical Lorenz curve, \textit{i.e.} $L\left(\frac{i}{n}\right)=\sum^{i}_{j=1}x_{j}/\sum^{n}_{j=1}x_{j}$, $i=1,2,\ldots,n$, superimposed by the theoretical curve $L_{\kappa}\left(u\right)$ given by Eq. \eqref{eq:Equation11} with estimates replacing $\alpha$ and $\kappa$ as necessary. This formula is shown by the solid line in the plots, and fits the data exceptionally well. The plots also exhibit a very good agreement between the empirical estimates of the Gini coefficient, obtained as $G=\frac{1}{n^{2}\mu}\sum^{n}_{i=1}\left(2i-n-1\right)x_{i}$, and the values returned by the analytical expression given by Eq. \eqref{eq:Equation12} for the estimated $\kappa$-generalized distribution; the 95\% confidence intervals constructed around the values of $G$ always cover the theoretical predictions $G_{\kappa}$\textsuperscript{6, 7}.
\footnotetext[6]{For the formulas used to estimate the empirical Lorenz curve and Gini coefficient see \textit{e.g.} Refs. \cite{Gastwirth1972} and \cite{Xu2003}, respectively.}\footnotetext[7]{The confidence intervals for the observed Gini coefficients have been calculated via the bootstrap resampling method based on 1000 replications. For general details about bootstrapping, see Refs. \cite{Bootstrap}.}
\par
In order to further evaluate the accuracy of our distributional model, we have also tested the hypothesis that each set of $n$ observed data follows a $\kappa$-generalized distribution by calculating the Kolmogorov-Smirnov (K-S) goodness-of-fit test statistic given by $D^{+}=\max_{1\leq i\leq n}\left[in^{-1}-P_{\leq}\left(x_{i}\right)\right]$, $i=1,2,\ldots,n$. Since in this case there is no asymptotic formula for calculating the $p$-value, we have reduced the problem to testing that the $x$ values have a standard exponential distribution (\textit{i.e.}, an exponential distribution with parameter equal to 1) by relating the function $P_{>}\left(x\right)$ given by Eq. \eqref{eq:Equation13} to the ordinary exponential function, namely $\exp_{\kappa}\left(-\beta x^{\alpha}\right)=\exp\left(-x_{\kappa}\right)$, through the transformation $x_{\kappa}=\frac{1}{\kappa}\log\left(\sqrt{1+\beta^{2}\kappa^{2}x^{2\alpha}}+\beta\kappa x^{\alpha}\right)$, where the parameters are estimated from the data. Thus the significance level in the upper tail is given approximatively by $P_{>}\left(T^{\ast}\right)=\exp\left[-2\left(T^{\ast}\right)^{2}\right]$, with $T^{\ast}=D^{+}\left(\sqrt{n}+0.12+0.11/\sqrt{n}\right)$, as suggested for example by Ref. \cite{Stephens1970}. The results are shown in the upper-left corner of panel \subref{fig:Figure1_d} of Figs. \ref{fig:Figure1} and \ref{fig:Figure2}. As one can appreciate, the maximum distance between the empirical data and the theoretical model as assessed by the K-S statistic is very small, and the $p$-values in parentheses do not lead to rejection of the null hypothesis that the data may come from a $\kappa$-generalized distribution at any of the usual significance levels (1\%, 5\% and 10\%). The linear behaviour emerging from the Quantile-Quantile (Q-Q) plots of the sample quantiles versus the corresponding quantiles of the fitted $\kappa$-generalized distribution displayed in the same panel strongly supports the quantitative results obtained by hypothesis testing.


\section{Summary and conclusions}
\label{sec:SummaryAndConclusions}
One of the main objectives of research on income distribution is to provide a mathematical description of the size distribution of income for approximating the underlying ``true'' distribution. Starting from Pareto contribution, a wide variety of functional forms have been considered as possible models for the distribution of personal income by size, and other approaches can no doubt be suggested and deserve attention.
\par
In this work we have proposed a three-parameter distribution by using a new approach having its root in the framework of the $\kappa$-generalized statistical mechanics. This model shows able to describe the entire income range, including the Pareto upper tail, and fits the Australian and US income data extremely well. The analysis of inequality performed in terms of its parameters reveals the merit of the proposed distribution, and provide the basis for a fruitful interaction between the two fields of statistical mechanics and economics.


\begin{ack}
T. Di Matteo wishes to thank the Australian Social Science Data Archive, ANU, for providing the ABS data and the partial support by ARC Discovery Projects: DP03440044 (2003) and DP0558183 (2005), COST P10 ``Physics of Risk'' project and M.I.U.R.-F.I.S.R. Project ``Ultra-high frequency dynamics of financial markets''.
\end{ack}


\end{document}